%% file: paper-bernoulli-poisson.tex
\documentclass[a4paper,10pt]{article}
\input{preambule-english}

\usepackage{srcltx}

\begin{document}

% -----------TITLE--------------

\title{Localization for random Schrödinger operators with low density potentials.}
\author{Francisco Hoecker-Escuti \thanks{The author wishes to thank his advisor Dr. Fr\'ed\'eric Klopp for having proposed this problem and the myriad of helpful discussions, the Centre Interfacultaire Bernoulli at the EPFL where part of this work was carried out and the ANR project ANR-08-BLAN-0261-01.}} 
\maketitle
\abstract{We prove that, for a density of disorder $\rho$ small enough, a certain class of discrete random Schrödinger operators on $\Z^d$ with diluted potentials exhibits a Lifschitz behaviour from the bottom of the spectrum up to energies at a distance of the order $\rho^\alpha$ from the bottom of the spectrum, with $\alpha>2(d+1)/d$. This leads to localization for the energies in this zone for these low density models. The same results hold for operators on the continuous, and in particular, with Bernoulli or Poisson random potential.}

\tableofcontents
% 
% \newpage
% \mbox{}
% \newpage
% 
% -----------TITLE--------------

\section{Introduction}

The purpose of this paper is to prove localization on an interval located at the bottom of the spectrum for some discrete and continuous  random models in a weak disorder regime, and a quantitative estimate on the size of this interval in terms of the disorder. This is achieved by showing a Lifschitz-like behaviour of the integrated density of states and, in the discrete case, the finite volume fractional moment criterion, whereas in the continuous, the initial step of the multi-scale analysis. Although the initial motivation was to study the Bernoulli--Anderson and the Poisson--Anderson model, in the discrete case we need to restrain ourselves to a certain class of diluted potentials, the random variables of which possess a regular distribution. Without this hypothesis, our main result concerning the exponential decay of the integrated density of states still applies.

By weak disorder we understand here that the mean potential is very small. This can be achieved, for example, by considering that the simple site potential is very small or that the disorder itself is very scarce. In the first case (and to which the terms weak disorder and weak localization are usually associated) it is very natural to multiply the potential in the Anderson model by a positive coupling constant $\lambda$
    \begin{displaymath}
      H_\omega=-\triangle + \lambda V_\omega
    \end{displaymath}
and study the behaviour for very small $\lambda$. There has been a number of works which establish localization (in chronological order: M. Aizenman \cite{aizenman1994localization}, W. Wang \cite{wang2001}, F. Klopp \cite{klopp2002a} \cite{klopp2002b}, A. Elgart \cite{elgart2008}) for this model in the weak coupling constant regime, in the discrete as well as in the continuous space. These results are obtained using the Frölich--Spencer multiscale analysis or the Aizenman--Molchanov fractional moment criteria. Lifschitz tails are a main ingredient and still the only mechanism understood to prove localization in dimensions greater than $2$.

 In this paper we consider low density disorder (or diluted) models. In these models, the impurities are large and rare rather thand small and dense. To fix the ideas, let us consider a protypical example. Let $H_\omega$ be a smoothed out version of the Bernoulli--Anderson model, defined by the Hamiltonian
  \begin{displaymath}
\label{def:anderson}
    H_\omega=H+V_\omega
  \end{displaymath}
  where $H$ is the free Laplacian on $\Z^d$ and $V_\omega$ the diagonal matrix defined by
  \begin{displaymath}
 	(V_\omega u)_n = \omega_n u_n \textrm{ for } u=(u_n)_{n\in \Z^d} \in \ell^2(\Z^d)
  \end{displaymath}
  with $(\omega_n)_{n\in \Z^d}$ independent identically distributed random variables with distribution
  \begin{displaymath}
	\Pro=(1-\rho)\delta_{0,\,\rho} + \rho \delta_{1,\,\rho},
  \end{displaymath}
  where $\delta_{\cdot,\rho}=\rho^{-1} v((x-\cdot)/\rho)$, $v$ being a positive mollifier ($v \geq 0, v \in C_0^\infty(\R)$, $\int_\R v(x)dx=1$, so $ \lim_{\rho \to 0} \delta_{\cdot,\rho} = \delta_\cdot$). Note that $\Esp(\omega_0) \sim \rho$.

    Under these assumptions we know that there exists a set $\Sigma_\rho \subset \R$ such that, for almost every $\omega$, the spectrum of the operator $H_\omega$ is equal to $\Sigma_\rho$. Moreover, if supp$(v)=[v_-,v_+]$, $\Sigma_\rho$ is given by
\begin{align}
 \Sigma_\rho = \sigma(- \triangle_{\Z^d}) + \textrm{supp} (\omega_0) = [-v_- \rho, 2d+ 1+ v_+ \rho]. \nonumber
\end{align}
By shifting the energy, we can assume that $\inf \Sigma_\rho=0$. This is no restriction as our purpose is to study the spectral properties of $H_\omega$ near the bottom of the spectrum and these remain unchanged.

We will prove the following theorem.
\begin{thm}
\label{thm:introlocal}
  Fix $\alpha>2(d+1)/d$ and $s \in (0, 1)$. There exists $\rho^*=\rho^*(\alpha,s)$ and $a>0$ such that for $ \rho \in (0,\rho^*)$, the Green's function of $H_\omega$ satisfies, for $(m,n) \in \Z^d \times \Z^d$ and for $E \in [0, \rho^\alpha]$,
  \begin{displaymath}
    \sup_{\varepsilon \in \R} \Esp \left[ \left| \left\langle \delta_n, \left( H_\omega - E -i\varepsilon \right)^{-1} \delta_m \right\rangle \right|^s \right] \leq \frac{1}{a} e^{-a \delta(E) |m-n|}
  \end{displaymath}
Here $\delta_n$ is the vector in $\ell^2(\Z^d)$ with all coordinates equal to $0$, except the $n$-th which is equal to $1$.
\end{thm}

The spectral consequences of this bound are well known \cite{aizenmanetal2001}, \cite{simon2006singular}, namely that we have that in the energy interval $[0,\rho^\alpha]$ this model exhibits exponential localization \cite{aizenman1993localization}, \cite{simon2006singular}, dynamical localization \cite{aizenman1994localization}, \cite{aizenman1998localization} and absence of level repulsion \cite{minami1996local}. These properties are detailed in \cite{klopp2002a}. 

  That Lifschitz tails are a hallmark of localization has been well known for physicists and mathematicians for long now \cite{lifshitz1964energy}, \cite{pastur297spectra}. In the weak disorder regime, it is expected to find Lifschitz-like behaviour in an interval going from the bottom of the spectrum up to a distance of the order of the variance from the mean. This leads to localization in this band, as shown by A. Elgart in \cite{elgart2008} for the discrete 3-dimensional model in a small coupling constant regime. The main difference with the low density regime is that here the variance is of the same order of the mean. In previous works \cite{klopp2002a}, \cite{klopp2002b} F. Klopp showed a similar result in a smaller band of the spectrum, through a scheme involving periodic approximations of the operator. This scheme have been proven quite robust, as it is used to handle the discrete and in the continuous model with no definite sign potential, and has been useful in other works. We use this scheme to prove the main results in this paper, but to get the best bound we needed to restrict ourselves to positive potentials. This restriction allow us to get better results, but it is not needed for the methods to work.

Theorem \ref{thm:introlocal} will be a consequence of an estimate on the \emph{integrated density of states}, which we define as:
  
  \begin{equation}
    \label{def:ids}
    \IDS(E)= \lim_{| \Lambda| \to +\infty} \frac{\# \{ \textrm{eigenvalues of } H_{\omega}|_\Lambda \leq E \} }{|\Lambda|}
  \end{equation}
  where $\Lambda$ denotes a cube of centre $0$, $|\Lambda|=\# \Lambda$ and $H_{\omega}|_\Lambda$ the Hamiltonian $H_{\omega}$ restricted to the cube $\Lambda$ with Dirichlet boundary conditions.
   The limit exists $\omega$--almost everywhere, it is non-random and non-decreasing \cite{carmona1990spectral}, \cite{pastur297spectra}.
Our main result in the discrete setting is:
\begin{thm}
\label{thm:introlifschitz}
	Let $\alpha > 2(d+1)/d$. Then there exists $\rho^*=\rho^*(\alpha) >0 $ and $\epsilon > 0$ such that for $\rho \in ]0,\rho^*[$, we have
	\begin{displaymath}
		\IDS( \rho^\alpha ) \leq e^{-\rho^{-\epsilon}}
	\end{displaymath}
\end{thm}

We now discuss the results on the continuous setting. We let $H_\omega$ 
defined as before
 \begin{align}
  H_\omega = H_0 + V_\omega
\end{align}
but here $H_0$ is the free Laplacian on $L^2(\R^d)$ and we let, for the Bernoulli--Anderson model,
\begin{align}
\label{def:randompotential}
  V_\omega(x) = \sum_{j \in \Z^d} \omega_j u(x-j),
\end{align}
where:
\begin{itemize}
 \item[HA] $\omega_j$ are independent, identically distributed Bernoulli random variables with probability $\varrho$.
\item[HB] $u \in L^\infty(\R^d,\R)$ is a compact supported simple-site potential and for $x \in \R$ we have
\begin{align}
  u_- \Id_{\Lambda_{\tau_-}(0)} \leq u(x) \leq u_+ \Id_{\Lambda_{\tau_+}(0)}
\end{align}
with $ 0< \tau_- < \tau_+$ and $0 < u_- < u_+$. The set $$\Lambda_{L}(x) = \{ x'=(x'_1,\ldots,x'_d) \in \R^d : -L-1/2 < x-x' \leq L + 1/2 \}$$ denotes the $d$-cube centered on $x$ and edge size $2L+1$.
\end{itemize}

Now let, for the Poisson--Anderson model,
\begin{align}
  V_\omega(x) = \sum_{\gamma \in \Gamma_\omega} u(x-y).
\end{align}
where:
\begin{itemize}
 \item[HC] $\Gamma_\omega$ is a Poisson process on $\R^d$ with density $\varrho>0$, i.e., for $A \subset \R^d$
\begin{align}
   \Pro \left( \# \{ \Gamma_\omega \cap B \} = k \right) = e^{-\varrho |B|} \left( \rho |B| \right)^k / k!
\end{align}
\end{itemize}
and $u$ as in (HB).

We define the integrated density of states as in \eqref{def:ids} (with $|\Lambda|$ now meaning the volume of the cube). Our main result in the continuous setting is:
\begin{thm}
\label{thm:introlifschitzcontinuous}
   The conclusion of Theorem \ref{thm:introlifschitz} is still valid for the Bernoulli--Anderson model under assumptions (HA)+(HB) and for the Poisson--Anderson model under assumptions (HB)+(HC).
\end{thm}
An inmediate consequence
will be the initial length scale estimate needed as input for the multiscale analysis.
This is shown in section \ref{sec:loccont}.
As previously commented, we are able to show localization in much more generality thanks to very recent progress  \cite{aizenmanetal2009}, \cite{bourgainkenig2005}, \cite{germinetetal2005}, \cite{germinetklein2011loc}. For a detailed discussion of the consequences of the mulstiscale analysis 
and the localization properties that follows, we refer the reader to Theorem 1.2(B) and Corollary 1.4
in \cite{germinetklein2011loc}.

\section{Discrete setting.}

\subsection{Assumptions.}
Let $\Hi=\ell^2(\Z^d)$ and $H:\Hi \to \Hi$ a translational invariant Jacobi matrix ---the Laplacian, for example--- with exponential off-diagonal-decay, i.e.
\begin{displaymath}
  H= \left( h_{k-k} \right)_{k,k' \in \Z^d}
\end{displaymath}
such that,
\begin{enumerate}
\item[H0] $h_{-k} = \overline{h_k}$; $k \in \Z^D$, and for some $k \neq 0$, $h_k \neq 0$
and there exists $c>0$ such that for $k \in \Z^d$ 
\begin{displaymath}
  \left| h_k \right| \leq \frac{1}{c} e^{-c |k|}.
\end{displaymath}
By Fourier transform
\begin{align}
\label{def:Fourier}
  \Fourier: l^2(\Z^d) \to L^2(\T^d)
\end{align}
where $\T^d = \R^d / (2 \pi \Z^d)$ we have
\begin{align}
  Hu=\Fourier^{-1} h \, \Fourier u \nonumber
\end{align}
where the diffusion law $h$ is real analytic on $\T^d$.
\end{enumerate}

We assume futhermore that
\begin{enumerate}
\item[H1] the minima of $h:\T^d \to \R^d$ are quadratic non-degenerate.
\end{enumerate}

Let $V^\omega$ be defined by
\begin{align}
  \left( V_\omega u \right)_n = \omega_n u_n \nonumber
\end{align}
for $u= \left( u_n \right)_{n \in \Z^d} \in l^2(\Z^d)$

\begin{enumerate} 
\item[H2] The random variables $\omega_n$ are independent, identically distributed, non trivial and bounded by $\omega_+$,. We assume furthermore that their essential infimum is $0$. There is no loss of generality as we may add a constant to the Hamiltonian without changing its spectral properties, as soon as the random variables are lower semibounded. Furthermore we assume that they satisfy
\begin{align}
  \Esp \left[ \omega_n \right] = \Esp \left[ \omega_0 \right] = \rho < \infty. \nonumber
\end{align}
\end{enumerate}

Our main result is
\begin{thm}
\label{thm:lifschitz}
	Assume (H1) and (H2). Let $\alpha > 2(d+1)/d$. Then there exists $\rho^*=\rho^*(\alpha) >0 $ and $\epsilon > 0$ such that for $\rho \in ]0,\rho^*[$, we have
	\begin{displaymath}
		\IDS( \rho^\alpha ) \leq e^{-\rho^{-\epsilon}}.
	\end{displaymath}
\end{thm}

Unfortunately, in the discrete case, a proof of localization for models with arbitrary random variables has yet to be proven. In order to use our results to get localization we need some regularity assumptions on the distribution of the random variables:

\begin{enumerate}

\item[H3] The common distribution $\Pro$ of $(\omega_n)$ is Hölder-continuous for $\rho \in [0,1]$, with the constant depending in the following fashion: There exists $ \tau \in ]0,1[ $ and $C>0$ such that, for $a<b$, one has,
\begin{displaymath}
  \Pro \left[ \left\{ \omega_0\in [a,b] \right\} \right] \leq C_H \left| b-a \right|^\tau \rho^{ - \tau }
\end{displaymath}

\begin{rem}
The motivation for this dependence on $\rho$ comes from the small coupling constant regime.
One may reinterpret this regime as a change of the probability distribution by a change of
random variables $\tilde \omega_n=\lambda \omega_n$. If one assumes $\tau$--Hölder
 continuity of the probability distribution, then the change of the Hölder constant
with respect to $\lambda$ take this form.
\end{rem}
\end{enumerate}

Our second result deals with the decay of the Green's function:

\begin{thm}
\label{thm:local}
  Assume (H1), (H2) and (H3). Fix $\alpha>2(d+1)$ and $s \in ]0, \tau/4[$. There exists $\rho^*=\rho^*(\alpha,s)$ and $a>0$ such that for $ \rho \in ]0,\rho^*[$, the Green's function satisfies, for $(m,n) \in \Z^d \times \Z^d$ and for $E \in [0, \rho^\alpha]$,
  \begin{displaymath}
    \sup_{\varepsilon \in \R} \Esp \left[ \left| \left\langle \delta_n, \left( H_\omega - E -i\varepsilon \right)^{-1} \delta_m \right\rangle \right|^s \right] \leq \frac{1}{a} e^{-a \delta(E) |m-n|}
  \end{displaymath}

\end{thm}

Theorems \ref{thm:introlocal} and \ref{thm:introlifschitz} are corollaries of Theorems \ref{thm:local} and \ref{thm:lifschitz} respectively.

\subsection{Localization}
\subsubsection*{Proof of Theorem \ref{thm:local}}

One way of showing localization from Lifschitz tails is to use the finite volume fractional moment localization criterion in \cite{aizenmanetal2001}.
Let $C_{0,L}$ be a cube in $\Z^d$ centered at $0$ and of sidelength $2L+1$. Let $H_\omega^D|_{C_{0,L}}$ be the random Hamiltonian $H_\omega$ restricted to the box $C_{0,L}$ with Dirichlet condition, i.e., $H_\omega^D|_{C_{0,L}} = \Pi_{C_{0,L}} H^D_\omega \Pi_{C_{0,L}}$.

Even though our model lacks a coupling constant (or it is equal to one), the small disorder parameter $\rho$ plays the same role and appears through the constants involved in the criterion. So the main difference with the calculation in \cite{klopp2002a} is that these constants may grow when $\rho$ gets small; they are nevertheless bounded by a polynomial in $\rho^{-s}$, $s \in ]0, \tau/4[$. This is because we have chosen the distribution to behave explicitly as in (H3) in function of $\rho$.
We recall from \cite{aizenmanetal2001} that, under assumptions (H2)--(H3), the following \emph{a priori} fractional moment bound
\begin{align}
\label{ineq:apriorifracmom}
  \Esp \left[ \left| \left\langle \delta_n, \left( H_\omega^D|_{C_{0,L}} - E -i\varepsilon \right)^{-1} \delta_m \right\rangle \right|^s \right] \leq C_s \rho^{-s} 
\end{align}
holds. Let us call for the sake of brevity
\begin{align}
 G^\omega_{mn} := \left\langle \delta_n, \left( H_\omega^D|_{C_{0,L}} - E -i\varepsilon \right)^{-1} \delta_m \right\rangle  \nonumber.
\end{align}
With our notation, we need to check that
\begin{displaymath}
  D L^{2d} \Xi(\rho^{-s}) 
   \sum\limits_{\substack{ m \in c_{0,L} \\ n \in \Z^d \backslash C_{0,L} } }
  e^{-c |m-n|} 
   \Esp \left[ \left| G^\omega_{mn} \right|^s \right] e^{\delta(E) |n|/D} < 1
\end{displaymath}
where $D$ is a constant depending on $h$ and the Hölder constant $C_H$, and $\Xi(\cdot)$ grows at most polynomially.

Define,
\begin{displaymath}
  \Omega_{\rho,\alpha,L} := \left\{ \textrm{there exists an eigenvalue of } H_\omega|_{C_{0,L}} \textrm{ in } [0,\rho^\alpha] \right\}.
\end{displaymath}

To check the finite volume fractional moment localization criterion, we will estimate the following expectation:
\begin{align}
\label{eq:fracmoment}
  \Esp \left[ \left| G^\omega_{mn} \right|^s \right] = \Esp \left[ \left| G^\omega_{mn} \right|^s \Id_{\Omega_{\rho,\alpha,L}}\right] + \Esp \left[ \left| G^\omega_{mn} \right|^s  \Id_{^c\Omega_{\rho,\alpha,L}} \right]
\end{align}

We proceed as follows: to estimate the first term we use the exponential bound for the integrated density of states we proved in Theorem \ref{thm:lifschitz} and for the second term we use a Combes--Thomas estimate. By using Hölder's inequality, the first term in \eqref{eq:fracmoment}, for fixed $0<s<s'<1$ and some $\epsilon>0$,
\begin{align}
\label{eq:firstholder}
   \Esp \left[ \left| G^\omega_{mn} \right|^s \Id_{\omega_{\rho,\alpha,L}} \right] \leq  \Esp \left[ \left| G^\omega_{mn} \right|^{s'} \right]^{s/s'} \Pro \left[ \Omega_{\rho, \alpha, L} \right]^{(s'-s)/s'}.
\end{align}
We will need the following theorem \cite{klopp2002a}, \cite{kloppwolff2002}:
\begin{thm}
  There exists $C>0$ such that, for $L \geq 1$, $\rho \in [0,1]$ and $E \in \R$ one has
  \begin{displaymath}
    \Pro \left[ \{ \left. H_\omega^D \right|_{C_{0,L}} \textrm{ admits an eigenvalue below } E \} \right] \leq CL^d \IDS(E).
  \end{displaymath}
\end{thm}

Let $\alpha>2(d+1)/d$. Our main result (Theorem \ref{thm:lifschitz}) together with the last theorem imply that there exists $\rho^*>0$ and $\epsilon>0$ such that for $0<\rho<\rho^*$, $1 \leq L \leq e^{\, \varrho^{-\epsilon/2}}$ one has
\begin{displaymath}
  \Pro\left[ \Omega_{\rho,\alpha,L} \right] \leq C e^{d\rho^{-\epsilon/2}}e^{-\rho^{-\epsilon}} \leq Ce^{-\frac{1}{2}\rho^{-\epsilon}}
\end{displaymath}
and now, using the \emph{a priori} estimation \eqref{ineq:apriorifracmom}, we conclude that \eqref{eq:firstholder} may be bounded by
\begin{displaymath}
  Ce^{-\frac{1}{2}\rho^{-\epsilon}}.
\end{displaymath}

Now, by a Combes--Thomas estimate (Lemma 6.1 in \cite{klopp2002a}), we get that, for $E \in [0,\rho^{\alpha'}]$, the second term in \eqref{eq:fracmoment} satisfies
\begin{displaymath}
  \Esp \left[ \left| G_{nm}^\omega \right|^s \Id_{\Omega{\rho,\alpha,L}} \right] \leq C \rho^\alpha e^{- \sqrt{|E-\rho^\alpha|} |m-n|/C}
\end{displaymath}
with $\alpha'>\alpha$.

Summing these bounds over $m\in C_{0,L}$ for $n \in \Z^d \backslash C_{0,L}$, and taking $1 \leq L \leq e^{\, \rho^{-\epsilon/2}}$, for $\rho$ small enough, we obtain:
\begin{align}
\label{eq:fracmomentest}
  &C L^{2d} \Xi(\rho^{-s}) \sum_{\substack{m \in C_{0,L} \\ n \in \Z^d \diagdown C_{0,L}}} e^{-c |m-n|} \Esp \left[ \left| G_{nm}^\omega \right|^s \right] e^{\delta(E)|n|/8C} \\
  & \leq C \Xi(\rho^{-s}) \left[ L^{3d}e^{\delta(E)L/8C} e^{-\rho^{-\epsilon}} + L^{2d}S\right] \nonumber
\end{align}
where
\begin{align}
\label{eq:fracmomentestS}
  S &:= \sum_{\substack{m \in C_{0,L} \\ n \in \Z^d \diagdown C_{0,L}}} e^{-c |m-n|} e^{-\delta(E)|m|/C} e^{\delta(E)|n|/8C}  \nonumber \\
  & \phantom{:} = \sum_{\substack{|m| \leq L \\ |n| \geq 2L}} + \sum_{\substack{|m| \leq L/2 \\ |n| \geq L}} \sum_{\substack{L/2 < |m| < L \\ L< |n| <2L}} e^{-c |m-n|} e^{-\delta(E)|m|/C} e^{\delta(E)|n|/8C} \nonumber \\
  & \phantom{:} \leq C e^{-L/C} L^{d-1} + CL^d e^{-\delta(E)/8C}.
\end{align}

If we take $\rho^{-\gamma} \leq L \leq e^{-\rho^{-\epsilon/2}}$ with $\gamma > \alpha/2$, then, for $E \in [0,\rho^{\alpha'}]$, one has $\delta(E) L \geq \rho^{-\epsilon}$ for some $\epsilon>0$ and $\rho$ sufficiently small. Hence using this in \eqref{eq:fracmomentest} and \eqref{eq:fracmomentestS}, for $\rho$ small enough, we obtain
\begin{displaymath}
  C L^{2d}\Xi(\rho^{-s}) \sum_{\substack{m \in C_{0,L} \\ n \in \Z^d \diagdown C_{0,L}}} e^{-c |m-n|} \Esp \left[ \left| G_{nm}^\omega \right|^s \right] e^{\delta(E)|n|/8C} < 1/16.
\end{displaymath}

So the finite volume criterion is satisfied if we take $C$ so that $8C>D$. Hence Theorem \ref{thm:lifschitz} implies Theorem \ref{thm:local}. 
\findem

We now turn to the proof of Theorem \ref{thm:lifschitz}.
\subsection{Klopp's Periodic Approximations}
Let $\omega \in \Omega$ and $N \in \N^*$. Define the periodic operator $H^N_\omega$ associated to 
\begin{align}
  H_\omega = H + V_\omega \nonumber
\end{align}
as
\begin{align}
  H_{\omega}^N=H+V^N_{\omega}=H+ \sum_{n \in \Z^d_{2N+1}} \omega_n \sum_{l \in (2N+1)\Z^d} |\delta_{l+n}\rangle \langle \delta_{l+n}| \nonumber
\end{align}
where $\Z^d_{2N+1}=\Z^d/(2N+1)\Z^d$. For the periodic operator, we define the integrated density of states (as in \eqref{def:ids}) and denote it by $\IDS_{\omega}^N$. The following lemma from \cite{klopp2002a} yields a very good approximation for the integrated density of states.

\begin{lem}
\label{lem:fastapprox}
	Let $\alpha>0$. There exists $\nu_0 \in (0,1)$ and $\gamma > 0$ such that, for $\rho \in [0,1]$, $E \in \R$, $\nu \in (0,\nu_0)$ and $N \geq \nu^{-\gamma}$ one has
	\begin{displaymath}
		\Esp(\IDS_{\omega}^N(E-\nu)) - e^{-\nu^{-\alpha}} \leq
		\IDS(E) \leq
		\Esp(\IDS_{\omega}^N(E+\nu)) + e^{-\nu^{-\alpha}}
	\end{displaymath}

\end{lem}

\subsection{Floquet Theory}
In this section we introduce some standard notions (see e.g.  \cite{kuchment1982floquet}, \cite{reedsimon04}). We follow the notations in \cite{klopp2002a}. The operator $H^N_\omega$ being periodic, we can use Floquet theory to reduce it to  an operator
acting on $$L^2 \left( \left[ - \frac{\pi}{2N+1}, \frac{\pi}{2N+1} \right]^d
\right)  \otimes \ell^2 \left( \Z^d_{2N+1} \right).$$ Define the unitary
transformation:
\begin{align}
  U: L^2 \left( \left[ - \pi, \pi \right]^d
  \right) \to L^2 \left( \left[ - \frac{\pi}{2N+1}, \frac{\pi}{2N+1} \right]^d
  \right)  \otimes \ell^2 \left( \Z^d_{2N+1} \right) \nonumber
\end{align}
by $(Uu)(\theta) = (u_k)(\theta)_{k\in \Z^d_{2N+1}}$; where the $(u_k(\theta))_{k\in \Z^d_{2N+1}}$ are defined by
\begin{align}
\label{eq:floqdecomp}
 u(\theta)= \sum_{k \in \Z^d_{2N+1}} e^{ik\theta} u_k(\theta)
\end{align}
and the functions $(\theta \mapsto u_k(\theta))_{k \in \Z^d_{2N+1}}$ are
$\frac{2\pi}{2N+1} \Z^d$-periodic.

Now the operator $U \Fourier H^N_\omega \Fourier^* U^*$ ---$\Fourier$ being
the Fourier transform \eqref{def:Fourier}--- is the multiplication by the matrix:
\begin{align}
 M_\omega^N(\theta) = H^N(\theta) + V_\omega^N \nonumber
\end{align}
where
\begin{align}
 H^N(\theta) = \big(\big( h_{j-j'}(\theta) \big)\big)_{(j,j') \in (\Z^d_{2N+1})^2}  \nonumber
\end{align}
and
\begin{align}
 V^N_\omega=\big(\big(\omega_j \delta_{jj'} \big)\big)_{(j,j') \in (\Z^d_{2N+1})^2}. \nonumber
\end{align}

Here, the functions $(h_k)_{k\in \Z^d_{2N+1}}$ are the components of $h$
decomposed according to \eqref{eq:floqdecomp}. The $(2N+1)^d \times (2N+1)^d$
matrices $H^N(\theta)$ and $V^N_\omega$ are non-negative.

Floquet theory gives us a useful characterization of $\IDS^N_\omega$ (see \cite{shubin1979spectral}):

  \begin{equation}
\label{eq:floquetIds}
    \IDS^N_\omega(E)=\frac{1}{(2\pi)^d} \int_{[-\frac{\pi}{2N+1},\frac{\pi}{2N+1}]^d} \# \{ \textrm{e.v. of } M^N_{\omega,p}(\theta) \textrm{ in } [0,E]\} ~d\theta.
  \end{equation}

Considering $H$ as $(2N+1)$-periodic on $\Z^d$, we see that the Floquet
eigenvalues of $H$ (for the quasi-momentum $\theta$) are $\left( h \left(\theta +
\frac{2 \pi k}{2N+1} \right) \right)_{k\in \Z^d_{2N+1}}$; the Floquet
eigenvalue $h \left(\theta + \frac{2 \pi k}{2N+1} \right)$ is associated to the
Floquet eigenvector $u_k(\theta)$, $k \in \Z^d_{2N+1}$ defined by
\begin{align}
 u_k(\theta) = \frac{1}{(2N+1)^{d/2}} \left( e^{-i \left(\theta +
\frac{2 \pi k}{2N+1} \right) j} \right)_{j \in \Z^d_{2N+1}}. \nonumber
\end{align}
In the sequel, the vectors in $l^2(\Z^d_{2N+1})$ are given by their components
in the orthonormal basis $( u_k(\theta) )_{k \in \Z^d_{2N+1}}$. The vectors of
the canonical basis denoted by $(v_l(\theta))_{l \in \Z^d_{2N+1}}$ have the
following components in this basis
\begin{align}
 v_l(\theta) = \frac{1}{(2N+1)^{d/2}} \left( e^{i \left(\theta +
\frac{2 \pi k}{2N+1} \right) l} \right)_{k \in \Z^d_{2N+1}}. \nonumber
\end{align}
We define the vectors $(v_l)_{l \in \Z^d_{2N+1}}$ by
\begin{align}
 v_l = e^{-i l \theta} v_l(\theta) = \frac{1}{(2N+1)^{d/2}} \left( e^{i
\left( \frac{2 \pi k}{2N+1} \right) j} \right)_{k \in \Z^d_{2N+1}}. \nonumber
\end{align}

\subsection*{Proof of Theorem \ref{thm:lifschitz}}

  As we have seen, the periodic approximation allows us to consider, $\Esp(\IDS_{\omega}^N(E))$  instead of $\IDS$ in order to show the scarcity of eigenvalues. By taking the expectation in \eqref{eq:floquetIds} (see \cite{klopp2002a} for more details), we get the following bound:
  \begin{align}
    \Esp(\IDS_{\omega}^N(E)) \leq C \Pro\left\{ \Omega(\rho^\alpha,\rho,N) \right\} \nonumber
  \end{align}
  where we define the event
  \begin{displaymath}
 	\Omega(E,\rho,N)=\left\{\omega:~\exists ~ \theta \in \R^d \textrm{ such that }
	M_{\omega}^N (\theta) \textrm{ has an e.v. in } [0,E] \right\}.
  \end{displaymath}
So in order to prove Theorem \ref{thm:lifschitz}, it suffices to prove the following:
\begin{prop}
\label{prop:main}
 Pick $\alpha> \alpha' (d+1)/d > 2 (d+1)/d$ and $\gamma$ given by 
Lemma \ref{lem:fastapprox}. There exists $\rho^*=\rho^*(\alpha,\gamma) > 0$ and $\epsilon>0$ such that for $\rho \in (0,\rho^*)$ we have
\begin{displaymath}
 	\Pro[\Omega(\rho^\alpha,\rho,N)] \leq e^{-\rho^{-\epsilon}}
\end{displaymath}
where
\begin{displaymath}
 	2N+1=[\rho^{(\alpha'-\alpha)/4}]_o [\rho^{-\alpha'/4}]_o [\rho^{-\gamma}]_o
\end{displaymath}

Here $[n]_o$ denotes the smallest odd integer greater than or equal to $n$.
\end{prop}

\subsection{Proof of Proposition \ref{prop:main}}
Pick $\alpha>2 \frac{d+1}{d}$, $\gamma$ as in Lemma \ref{lem:fastapprox}, and let $\frac{d}{d+1} \alpha>\alpha'>2$.
By (H0), $h$ is real analytic on $\T^d$. Let $Z$ be the finite set of minima of $h$
\begin{align}
  Z=\{\theta_1, \ldots, \theta_M\}. \nonumber
\end{align}

By (H1), we know that there exists $C>0$ such that, for $\theta \in \T^d$
\begin{align}
\label{ineq:quad}
  h(\theta) \geq C \min_{1 \leq J \leq M} |\theta - \theta_J |^2.
\end{align}
$C$ is a constant that may change from line to line.

Let
\begin{displaymath}
 2L+1=[\rho^{(\alpha'-\alpha)/2}]_o [\rho^{-\alpha'/4}]_o~~\textrm{,}~~2K+1= [\rho^{-\gamma}]_o
\end{displaymath}
and $\omega \in \Omega(\rho^\alpha,\rho,N)$. Note that $2N+1=(2L+1)(2K+1)$. Hence, there exists $\theta \in \R^d$ and $a=\sum a_k u_k(\theta)$ such that
\begin{itemize}
 \item $\|a\|_{l^2(\Z^d_{2N+1})} = \sqrt{\sum_{k \in \Z^d_{2N+1}} |a_k|^2}=1$
 \item $\langle M^N_\omega(\theta) a,a \rangle_{l^2(\Z^d_{2N+1})} \leq \rho^\alpha$
\end{itemize}

As the operators $H^N(\theta)$ and $V^N_\omega$ are non negative, one gets:
\begin{equation}
\label{ineq:1}
 \langle H^N(\theta) a,a \rangle_{l^2(\Z^d_{2N+1})} \leq \rho^\alpha 
\end{equation}
and
\begin{equation}
\label{ineq:2}
 \langle V^N_\omega a,a \rangle_{l^2(\Z^d_{2N+1})} \leq \rho^\alpha.
\end{equation}

By \eqref{ineq:quad}, we know that, for $1 \leq J \leq M$, $\theta \in [\frac{-\pi}{2N+1},\frac{\pi}{2N+1}]^d$, some $C>0$ and $\rho$ small enough, one has

\begin{equation}
\label{ineq:3}
 \left| \frac{2\pi k}{2N+1} - \theta_J \right| \geq \frac{1}{2L+1} \Longrightarrow \left( h \left( \theta + \frac{2\pi k}{2N+1} \right) \geq \rho^{\alpha-\alpha'/2} /C \right).
\end{equation}

For $1 \leq J \leq M$, let $k_J \in \Z^d$ be the unique vector satisfying $$2\pi k_J - (2N+1)\theta_m \in [-\pi,\pi)^d$$ and let 
\begin{displaymath}
 (a^J)_k = \left\{ \begin{array}{ll} a_k & \textrm{if } |k-k_J| \leq K \\ 0 & \textrm{ if not}
\end{array} \right.
\end{displaymath}

For $\rho$ sufficiently small, the vectors $\left( a^J \right)$ are pairwise orthogonal. By \eqref{ineq:1} and \eqref{ineq:3}, we have that
\begin{align}
\label{ineq:4}
 \left\| a - \sum_{J=1}^M a^J \right\|_{l^2(\Z^d_{2N+1})} \leq C\rho^{\alpha'/4} 
\end{align}

Now we write
\begin{align}
\label{eq:sum}
    \left\langle V^N_\omega a, a \right\rangle 
    = & \underbrace{ \left\langle V^N_\omega \left( \sum_{J=1}^M a^J \right), \left( \sum_{J=1}^M a^J \right) \right\rangle }_{(i)} \notag \\ 
  + 2 \textrm{Re} & \underbrace{ \left\langle V^N_\omega \left( \sum_{J=1}^M a^J \right), \left( a - \sum_{J=1}^M a^J \right) \right\rangle }_{(ii)} \notag \\ 
  + & \underbrace{ \left\langle V^N_\omega  \left( a - \sum_{J=1}^M a^J \right), \left( a - \sum_{J=1}^M a^J \right) \right\rangle }_{(iii)}.
\end{align}

Using \eqref{ineq:4}, the third term $(iii)$ in the sum satisfies, for $\rho$ small enough,
\begin{align}
  \left| \left\langle V^N_\omega  \left( a - \sum_{J=1}^M a^J \right) , \left( a - \sum_{J=1}^M a^J \right) \right\rangle \right| 
  & \leq \left\| a - \sum_{J=1}^M a^J \right\|_2^2 \nonumber
   \\  &\leq C \rho^{\alpha'/2}.\nonumber
\end{align}

Now assume for a moment that the second term $(ii)$ in the sum \eqref{eq:sum} satisfies
\begin{align}
\label{ineq:hypXter}
  \rho^{\frac{3}{4}+\frac{\alpha'}{8}} < 
  \left| 2 \textrm{Re} \left\langle V^N_\omega \left( \sum_{J=1}^M a^J \right), \left( a - \sum_{J=1}^M a^J \right) \right\rangle \right|.
\end{align}
Since, by Cauchy--Schwarz
\begin{align}
  &\left| 2 \, \textrm{Re} \left\langle V^N_\omega \left( \sum_{J=1}^M a^J \right), \left( a - \sum_{J=1}^M a^J \right) \right\rangle \right| & \nonumber
  \\ &\leq 2 \sqrt{\left\langle V^N_\omega \left( \sum_{J=1}^M a^J \right), \left( \sum_{J=1}^M a^J \right) \right\rangle} \left\| a - \sum_{J=1}^M a^J \right\|_2,& \nonumber
\end{align}
we have that
\begin{align}
\label{ineq:Xter}
  \left| \left\langle V^N_\omega \left( \sum_{J=1}^M a^J \right), \left( \sum_{J=1}^M a^J \right) \right\rangle \right| \geq \frac{C}{2} \rho^{(6-\alpha')/{4}}, 
\end{align}
but the probability that this term is of the order $\rho^{1-\epsilon}$ is exponentially small, see Remark \ref{rem:case2} later on. Note that $3/4 + \alpha'/8>1$ and $ (6-\alpha')/{4} < 1$.

On the other hand, if \eqref{ineq:hypXter} is not true, in order to satisfy \eqref{ineq:2}, we must have, for $\rho$ small enough,
\begin{align}
  \label{ineq:6}
    \left| \left\langle V^N_\omega  \left( \sum_{J=1}^M a^J \right), \left( \sum_{J=1}^M a^J \right) \right\rangle \right| \leq \rho^{(\alpha'+6)/8}=\rho^{1+\epsilon}
\end{align}
as this is the order of the largest term (note that $1< (\alpha' + 6)/8< \alpha'/2$).

We will show that this happens with an exponentially small probability. To do so, we will need the following lemma,
	\begin{lem}[\cite{klopp2002a}]
	\label{lem:applemma}
	  Assume $N, L, K, L', K'$ positive integers such that:
	  \begin{itemize}
	    \item $2N+1 = (2L+1)(2K+1) =(2L'+1)(2K'+1)$,
	    \item $K<K'$ and $L'<L$.
	  \end{itemize}
	  For $a \in l^2(\Z_{2N+1})$ such that supp $a \subset C_{0,K}$, there exists $\tilde{a} \in l^2(\Z_{2N+1})$ with the following properties:
	  \begin{itemize}
	    \item we have that $\| a- \tilde{a} \|_{l^2(\Z_{2N+1})} \leq C_{K,K'} \| a \|_{l^2(\Z_{2N+1})} $ with $C_{K,K'} \asymp K/K'$,
	    \item the vector $\tilde{a}$ is constant over cubes $C_{\gamma,L}$ with $\gamma \in (2K+1)\Z^d$,
	    \item we have $\|a\|_{l^2(\Z_{2N+1})} =\| \tilde{a}\|_{l^2(\Z_{2N+1})} $.
	  \end{itemize}

	\end{lem}
Define
\begin{align}
   2L'+1=[\rho^{(\alpha'-\alpha)/2}]_o ~~\textrm{and}~~2K'+1=[\rho^{-\alpha'/4}]_o [\rho^{-\gamma}]_o. \nonumber
\end{align}

We now translate each of the $a^J$ by $k_J$ so as to centre their support at $0$. The vector obtained is denoted again by $a^J$. This allows us now to apply the lemma to each $a^J$, as $K/K' \sim \rho^{\alpha'/4}$, we have $\| a^J - \tilde{a}^J \|^2_{l^2(\Z_{2N+1})} \leq \rho^{\alpha'/2}$. Now we write,
\begin{align}
    \left\langle V^N_\omega \left( \sum_{J=1}^M a^J \right), \left( \sum_{J=1}^M a^J \right) \right\rangle 
    = & \underbrace{ \left\langle V^N_\omega \left( \sum_{J=1}^M \tilde{a}^J \right), \left( \sum_{J=1}^M \tilde{a}^J \right) \right\rangle  }_{(I)} \notag \\ \nonumber
  + 2 \textrm{Re} & \underbrace{ \left\langle V^N_\omega \left( \sum_{J=1}^M \tilde{a}^J \right), \left( \sum_{J=1}^M a^J - \tilde{a}^J \right) \right\rangle }_{(II)} \\ 
  + & \underbrace{ \left\langle V^N_\omega \left( \sum_{J=1}^M a^J - \tilde{a}^J \right), \left( \sum_{J=1}^M a^J - \tilde{a}^J \right) \right\rangle }_{(III)}.
\label{eq:sum2}
\end{align}

By Lemma \ref{lem:applemma}, the third term in this sum is bounded by $C M\rho^{\alpha'/2}$. Now, repeating the same trick as before, should the absolute value of the second term  $|(II)|$ be greater than $\rho^{\frac{3}{4}+\frac{\alpha'}{8}}$, we would have, by Cauchy--Schwarz,
\begin{align}
  \left\langle V^N_\omega \left( \sum_{J=1}^M \tilde{a}^J \right), \left( \sum_{J=1}^M \tilde{a}^J \right) \right\rangle \geq C\rho^{\frac{3}{2}-\frac{\alpha'}{4}}. \nonumber
\end{align}
On the other hand, if the condition $|(II)|>\rho^{\frac{3}{4}+\frac{\alpha'}{8}}$ is not fulfilled, the first term must be smaller than $C\rho^{(\alpha'+6)/8}$ for some constant $C>0$ and $\rho$ small enough. We thus conclude that there exists $C>0$ and at least one pair $J,J'$ for which either
\begin{align}
    \left\langle V^N_\omega \tilde{a}^J, \tilde{a}^{J'} \right\rangle \leq C\rho^{\frac{\alpha'}{8}+\frac{3}{4}} \nonumber
\end{align}
or
\begin{align}
    \left\langle V^N_\omega \tilde{a}^J, \tilde{a}^{J'} \right\rangle \geq C\rho^{\frac{3}{2}-\frac{\alpha'}{4}} \nonumber
\end{align}
for $\rho$ small enough. 
These implies the two conditions
\begin{align}
    \pm \left\langle V^N_\omega \tilde{a}^J, \tilde{a}^{J'} \right\rangle \leq \pm C\rho^{1 \pm \epsilon} \nonumber
\end{align}
with $\epsilon=(\alpha' - 2)/8$.
\begin{rem}
\label{rem:case2}
We show by the same method that if \eqref{ineq:hypXter} holds, then \eqref{ineq:Xter} leads to the last inequality. Indeed, $(III)$ is always $\lesssim \rho^{1+\epsilon}$ and we saw that if $(II)$ is not $\lesssim \rho^{1+\epsilon}$ it lead to one of the last inequalities. By assuming \eqref{ineq:hypXter} we must then have $(I)\gtrsim \rho^{1-\epsilon}$.
\end{rem}

Remembering that we have translated the $a^J$ by $k_J$, we expand
\begin{align}
\nonumber
  \left\langle V^N_\omega \tilde{a}^J, \tilde{a}^{J'} \right\rangle  &= \sum_{l \in \Z^d_{2N+1}} e^{\frac{2 i \pi \left( k_J - k_{J'} \right) l}{2N+1}} \omega_l \left\langle \tilde{a}^J, v_l \right\rangle \overline{ \left\langle \tilde{a}^{J'}, v_l \right\rangle } \\ \nonumber
  &= \sum_{k' \in \Z^d_{2K'+1}} S(J,J',k) e^{\frac{2 i \pi \left( k_J - k_{J'} \right) k'}{2K'+1}} \left( 2L'+1 \right)^d 
  \\& \phantom{= \sum_{k' \in \Z^d_{2K'+1}} S(J,J',k)} \times \langle \tilde{a}^J, v_l \rangle \overline{\left\langle \tilde{a}^{J'}, v_l \right\rangle } \notag
\end{align}
where 
\begin{align}
  S(J,J',k') = \frac{1}{\left( 2L'+1 \right)^d } \sum_{l' \in \Z^d_{2L'+1}} \omega_{l' + k'(2L'+1)}  e^{\frac{2 i \pi \left( k_J - k_{J'} \right) l'}{2L'+1}}. \nonumber
\end{align}
If we define
\begin{align}
  \Sigma(J,J',k') = \frac{1}{\left( 2L'+1 \right)^d } \sum_{l' \in \Z^d_{2L'+1}} \omega_{l' + k'(2L'+1)} e^{i \left( \theta_J - \theta_{J'} \right) l'} \nonumber
\end{align}
we note that
\begin{align}
  \left| \Sigma(J,J',k') - S(J,J',k') \right| = O(\rho^\gamma) \nonumber
\end{align}
since $ \left| \frac{2 i \pi \left( k_J - k_{J'} \right) l'}{2L'+1} - \theta_J \right| \leq \frac{1}{2N+1}$. As $\| \tilde{a}_J\|=\|a_J\|\leq 2$ we get that
\begin{align}
  \pm \sum_{k' \in \Z^d_{2K'+1}} \Sigma(J,J',k) e^{\frac{2 i \pi \left( k_J - k_{J'} \right) k'}{2K'+1}} \left( 2L'+1 \right)^d \nonumber \\
  \\ \phantom{= \sum_{k' \in \Z^d_{2K'+1}} S(J,J',k)} \times \langle \tilde{a}^J, v_l \rangle \overline{\left\langle \tilde{a}^{J'}, v_l \right\rangle } \leq \pm C\rho^{1\pm\epsilon}
\end{align}
and we conclude that if $\omega \in \Omega(\rho,\rho^{\alpha},N)$ then for some $1 \leq J \leq J' \leq M$ and $k' \in \Z^d_{2K'+1}$, we have	
\begin{align}
  \nonumber \pm \left| \frac{1}{\left( 2L'+1 \right)^d} \sum_{l' \in \Z^d_{2L'+1}} \omega_{l' + k'(2L'+1)} e^{i (\theta_J - \theta_{J'})l'} \right| \leq \pm C\rho^{1 \pm \epsilon}.
\end{align}

By a reduction similar to the one found in the proof of Proposition 4.2 in \cite{klopp2002a}, we can get rid of the exponential terms in the left-hand side. We summarize what we have obtained in the following lemma.
\begin{lem}
 Pick $\alpha > \alpha' > 2$ and $N$ as in the Proposition. Let $L'$ and $K'$ defined as before. There exists $C>0$ and $\rho_0$ such that for $0 < \rho < \rho_0$ we have
\begin{displaymath}
 \Omega(\rho^\alpha,\rho,N) \subset \bigcup_{|k'| \leq K'} \left( \bigcup_{1 \leq J \leq J' \leq M} \Omega^{J,J',k'}_+ \cup \Omega^{J,J',k'}_- \right)
\end{displaymath}
where for $1 \leq J \leq J' \leq M$ and $|k'| \leq K$ we define
\begin{displaymath}
 \Omega^{J,J',k'}_\pm= \left\{ \omega:~ \pm \frac{1}{(2L'+1)^d} \sum_{|l'| \leq L'} \omega_{k'(2L+1)+l'} \leq \pm C \rho^{1\pm\epsilon} \right\}
\end{displaymath}
\end{lem}

\mbox{} \newline
If there exists $\epsilon>0$ such that, for $\rho$ sufficiently small,
\begin{align}
  \nonumber \Pro\left\{ \Omega^{J,J',k'}_\pm \right\} \leq e^{-\rho^{-\epsilon}}.
\end{align}
the theorem is proven as the number of sets in the union in the last lemma is bounded by $\rho^{-1}$. This means that we need to prove that the following probabilities:
\begin{displaymath}
 \Pro \left( \frac{1}{(2L'+1)^d} \sum_{|l'| \leq L'} \omega_{k'(2L+1)+l'} \leq C \rho^{1+\epsilon} \right)
\end{displaymath}
and
\begin{displaymath}
 \Pro \left( \frac{1}{(2L'+1)^d} \sum_{|l'| \leq L'} \omega_{k'(2L+1)+l'} \geq C \rho^{1-\epsilon} \right)
\end{displaymath}
are exponentially small. This can be done using classical large deviation theory. We will do it succinctly for one of the inequalities. We reindex the random variables as $\omega_U$, $U=1,\ldots,R=(2L'+1)^d$; then use Markov's inequality to obtain:
\begin{align}
\label{ineq:largedev}
  \Pro\left( \frac{1}{R} \sum_{U=1}^R \omega_U \leq C \rho^{1 + \epsilon} \right) \leq \Esp\left( e^{- t \sum \omega_U} \right) e^{CRt\rho^{1 + \epsilon}} \\
  = \prod_{U=1}^R \Esp\left( e^{- t \omega_0} \right) e^{ CRt\rho^{1 + \epsilon}},
  \nonumber
\end{align}
where we have used the fact that the random variables are independent, identically distributed.

Now, as long as $t \omega_+ < 1$, we get that there is a $C$ such that $\exp(- t \omega_0) < 1 - t\omega_0$ and thus
\begin{align}
  \nonumber \Esp\left( e^{-t \omega_0} \right) &< 1 - C t \Esp(\omega_0) \\
  &= 1 - C t \rho \leq e ^{-C t \rho}.
\end{align}
Note that we have used (H2).
Plugging this into \eqref{ineq:largedev}, there exists a $C$ such that,
\begin{align}
\nonumber
  \Pro\left( \frac{ 1}{R} \sum_{U=1}^R \omega_U \leq C \rho^{1+\epsilon} \right) \leq e^{-CRt(\rho + \rho^{1+\epsilon})} \leq e^{-\frac{1}{2} CR\rho}.
\end{align}
Noting now that, as $R \sim \rho^{d(\alpha' - \alpha)/2}$, and by hypothesis $d(\alpha' - \alpha)/2> \alpha'/2 > 1$, this probability is exponentially decaying. This proves the proposition.
\findem
\section{Continuous setting.}

\subsection{Assumptions.}

\label{sec:assumptions}
We start by setting our hypotheses in the continuous setting. Define a \emph{normalized Anderson Hamiltonian} $H_\omega$ as in
\eqref{def:anderson} in the introduction but we assume from now on:
\begin{itemize}
\item[(HD)] The operator $H_0 := -\triangle_{\R^d}+V_{per}$ where $\triangle_{\R^d}$ denotes
 the free Laplacian on $\R^d$ and $V_{per}$ is a bounded $q\Z^d$-periodic potential with $q=(2\hat q +1) >1$, an integer which we take odd for convenience sake. We assume furthermore that $H_0$ has the unique continuation principle (UCP), that is, for any $E \in \R$ and for any function $\phi \in H^2_{\textrm{loc}} (\R^d)$, if 
$(H_0 -E)\phi=0$, and if $\phi$ vanishes on an open set, then $\phi \equiv 0$.
\end{itemize}
The UCP has been used to obtain Wegner estimates (as in \cite{combesetal2007wegner}, \cite{combesetal2003}) and it is in particular verified under our hypotheses for $d\geq 3$ (\cite{wolff1995ucp}). 
\begin{itemize}
\item[(HE)] The potential $V_\omega$ is defined as in \eqref{def:randompotential} in the introduction but we let
  $\omega_n$ be non degenerate, independent and identically distributed random
  variables satisfying $\{0,1\}\in \supp \omega_0 \subset [0,1]$ and $\Esp \left[ \omega_0 \right] = \varrho < \infty$.
\end{itemize}
We would like to stress that (HD) is not really restrictive (see section 2 in \cite{germinetklein2011loc}). (HE) the analog of (H2) in the discrete case, 
but we will not need any regularity of the random variables distribution (as in (H3)).

From now on we will refer to the operator $H_\omega$ together with (HD), (HE), as \emph{normalized Anderson Hamiltonian} and  $H_\omega$ together with (HB), (HC), as \emph{Poisson--Anderson Hamiltonian}.

 The purpose of this section is to proof the following:
 \begin{thm}
\label{thm:lifschitzcontinuous}
  Assume (HB)+(HC) or (HB)+(HD)+(HE). Fix $\alpha > 2(d+1)/d$. There exists $\varrho^*=\varrho^*(\alpha) >0 $ and $\gamma >0$ such that, for $\varrho \in (0,\varrho^*)$, we have
 \begin{equation}
\label{eq:lifschitzcontinuous}
  	\IDS( \varrho^\alpha ) \leq e^{-\varrho^{-\epsilon}}.
 \end{equation}
 \end{thm}
Theorem \ref{thm:introlifschitzcontinuous} is just a corollary of \ref{thm:lifschitzcontinuous}.

\subsection{Localization}
\label{sec:loccont}
As discussed previously, exponential and dynamical localization are a consequence of the multiscale analysis with a Wegner estimate developed by Bourgain and Kenig in \cite{bourgainkenig2005} for the Bernoulli--Anderson model, and by Germinet, Hislop and Klein in \cite{germinetetal2005} for the model with Poisson potential. Being an induction procedure, we only need to check that some 'a priori' finite volume estimates holds. In order to use the results of these works, we need to be able to provide a number of 'free sites' with the initial length scale estimate. First we proceed with the normalized Anderson model.
\subsubsection*{Free sites}
We follow the proof of Theorem 4.3 in \cite{germinetklein2011loc}.
Given a box $\Lambda=\Lambda_L(x)$ in $\R^d$, we denote by $\tilde \Lambda$ the set $\Lambda \cap \Z^d$. Given $S \subset \tilde \Lambda$, $\mathbf{t}_S = \{ {t}_\zeta \}_{\zeta \in S} \in [0,1]^{S}$, set
\begin{align}
 H_{\omega,\mathbf{t}_S,\Lambda} := - \triangle_{\Lambda} + V_{per,\Lambda} + V_{\omega,\mathbf{t}_S,\Lambda} \quad \textrm{on} \quad L^2(\Lambda) \nonumber
\end{align}
where $\triangle_{\Lambda}$ is the restricted Laplacian with Dirichlet boundary conditions, $V_{per,\Lambda}$ is the restriction of $V_{per}$ to $\Lambda_L$ and
\begin{align}
 V_{\omega,\mathbf{t}_S,\Lambda} := \chi_\Lambda V_{\omega_\Lambda,\mathbf{t}_S} \nonumber
\end{align}
with
\begin{align}
 V_{\omega_\Lambda,\mathbf{t}_S}(x) :&= V{\omega_{\Lambda \slash \mathbf{t}_S}}(x) + V_{t_S}(x) \\
  &= \sum_{\zeta \in \tilde \Lambda \slash S} \omega_\zeta u_\zeta (x-\zeta) + \sum_{\zeta \in S} t_\zeta u_\zeta (x-\zeta). \nonumber
\end{align}
We need to show that the probability that the operator $H_{\omega,\mathbf{t}_S,\Lambda}$ has an eigenvalue under $\varrho^\alpha$ is exponentially small  and that this
happens uniformly with respect to $\mathbf{t}_S \in [0,1]^{S}$, for $S$ dense enough (see \cite{germinetklein2011loc}).

Set $\tilde q =\max\{3,q\}$, with $q$ as in (HD). For a given a box $\Lambda=\Lambda_L(x)$ in $\R^d$ we let
$$H^{(\tilde q)}_\omega := H_0 + V^{(\tilde q)}_\omega \quad \textrm{with} \quad V^{(\tilde q)}_\omega:= \sum_{\zeta \in \tilde q \Z^d} \omega_\zeta u(x-\zeta),$$
which is a normalized Anderson Hamiltonian for which the underlying lattice is $\tilde q \Z^d$ instead of $\Z^d$ and so its integrated density of states $\IDS^{(\tilde q)}(E)$ is well defined. We will only consider scales $L \in \tilde q \N$. Let
$$H^{(\tilde q)}_{\omega,\Lambda_L} := - \triangle_{\Lambda} + V_{per,\Lambda} + V^{(\tilde q)}_{\omega,\Lambda} \quad \textrm{on} \quad L^2(\Lambda)$$
where $V^{(\tilde q)}_{\omega,\Lambda}$ is the restriction of $V^{(\tilde q)}_{\omega}$ to $\Lambda_L$. We clearly have that, for any  $\mathbf{t}_S \in [0,1]^{S}$,
\begin{align}
\label{ineq:star}
H_{\omega,\mathbf{t}_S,\Lambda} \geq -\triangle_\Lambda +V_{per,\Lambda} + V_{\omega_{\Lambda \slash S}}.
\end{align}
Finally, define the (non-normalized) counting function
$$ N^{(\tilde q)}_{\omega,\Lambda_L}(E):= \textrm{tr}\,\, \chi_{]-\infty,E]} \left( \tilde H^{(\tilde q)}_{\omega,\Lambda_L} \right).$$

Setting $S= \tilde \Lambda_L(x) \backslash \tilde q \Z^d$, we claim that there exists $\epsilon>0$ such that,
$$ \Pro \left\{ H_{\omega,\mathbf{t}_S,\Lambda} \geq \varrho^\alpha \textrm{ for all } \mathbf{t}_{ S} \in [0,1]^{S} \right\} \geq 1 - e^{-\varrho^{-\epsilon}}$$
 for $\varrho$ small enough. To prove this, we remark first that the conclusion of Theorem \ref{thm:lifschitzcontinuous} is valid for $H^{(\tilde q)}_\omega$ (by changing the constants) and we remind that (see (VI.15) in \cite{carmona1990spectral}), 
$$ \mathbb E \left( N^{(\tilde q)}_{\omega,\Lambda_L}(E) \right) \leq \IDS^{(\tilde q)}(E) \left| \Lambda_L \right|, $$
and thus calling  $\Omega:=\{\omega:{H}^{(\tilde q)}_{\omega,\Lambda} \textrm{ has an e.v. in } [0,\varrho^\alpha] \}$ 
and using \eqref{eq:lifschitzcontinuous} and Markov's inequality we see that indeed
\begin{align}
 \Pro\left( \Omega \right) \leq e^{-\varrho^{-\epsilon}} e^{\, \varrho^{-\epsilon/2}} \leq e^{- \varrho^{-\epsilon/2}}
\end{align}
for $|\Lambda_L| \leq e^{\, \varrho^{-\epsilon}/2}$ and $L \in \tilde q \N$; so, by \eqref{ineq:star}, we get that, uniformly in the $\mathbf t_S$
\begin{align}
 \Pro \left( H_{\omega,t_S,\Lambda} \geq \varrho^\alpha \right) \leq 1 - e^{- \varrho^{-\epsilon}/2.}
\end{align}
As shown  in \cite{germinetklein2011loc}, this is also true for any $L$ in this range. This range of scales is enough to start the mulstiscale analysis (see Proposition 4.6 in \cite{germinetklein2011loc}). 

\subsubsection*{Poisson--Anderson model}

The existence of localization for the Poisson--Anderson Hamiltonian is a consequence of the same phenomenon, 
namely that with very good probability the effect of the random potential on finite volume operators
is to ``push'' the spectrum away from zero, uniformly with respect to free sites (suitably defined
for this model). We will explain briefly what is needed to proof, taking notation and definitions from \cite{germinetetal2005}.
 We will show that for $E \in [0,\varrho^{\alpha'}]$
the scales $\varrho^{-\epsilon} \lesssim |\Lambda| \lesssim e^{\varrho^{-\epsilon/2}}$ are $E$-localizing (see
definition 3.16 in \cite{germinetetal2005}), for a fixed $\alpha'>\alpha$ and $\varrho$ small enough.

The idea is the following. We start by subdividing a big cube $\Lambda=\Lambda_L$ in $\R^d$ 
in non overlapping cubes $\Lambda(j)$ of side $\eta:=e^{-L^{10^6 d}}$, indexed by:
$$\mathbb J_\Lambda := \{ j \in x+\eta \mathbb Z^d : \Lambda(j) \subset \Lambda \}$$
and, with very little cost in probability, we only need to consider configurations $X$
such that the number of points in $\Lambda$ are $\lesssim \varrho L^d$ and  at most
there is one point in each $\Lambda(j)$, i.e.
$$N_X \left( \Lambda \right) \lesssim \varrho L^d, \quad N_X \left( \Lambda(j) \right) \leq 1.$$
Here $N_X(\Lambda)$ is the random variable giving the number of points the configuration
$X$ puts in $\Lambda$.
These configurations are thus in bijection with
$$\mathcal J_\Lambda := \{ J \subset \mathbb J_\Lambda : \# J \lesssim \varrho L^d \}.$$
The next crucial observation by Germinet, Hislop and Klein is that we only need to
consider the configurations having their points centered in each $\Lambda(j)$. We can
indeed 'wiggle' the points inside each box $\Lambda(j)$ and by doing so move the
eigenvalues by no more than $\lesssim e^{-L^{1-\epsilon}}$. They introduced then an equivalence
relation (eq. (3.29)) in the space of configurations, the equivalence classes of which are
then indexed by $\mathcal J_\Lambda$. We write $[J]_\Lambda$  for the equivalence
class of the configuration having a point in the center of $\Lambda(j)$ whenever
$j \in J$ and $[J]_\Lambda \sqcup [J']_\Lambda$ for the disjoint union.

We define now the 'basic events' which take care of the free sites.
For a given set $B$, let $\mathcal P_0(B)$ the collection of its countable subsets. Given two 
configurations $X,Y \in \mathcal P_0(\R^d)$ and $\mathbf t_Y=\{t_\zeta\}_{\zeta \in Y} \in [0,1]^Y$ define
$H_{X,(Y,t_Y),\Lambda}$ as in equation (3.10) in \cite{germinetetal2005}:
$$H_{X,(Y,t_Y),\Lambda}:= - \triangle_\Lambda +  V_{X,(Y,t_Y),\Lambda} \quad \textrm{where} \quad V_{X,(Y,t_Y),\Lambda}:= \chi_\Lambda V_{X_\Lambda,(Y_\Lambda,t_{Y_\Lambda})}$$
and
$$V_{X,(Y,t_{Y})}:= V_X(x) + \sum_{\zeta \in Y} t_\zeta u(x-\zeta).$$

Let us recall that a Poisson process $\Upsilon_\omega$ with density $2\varrho$ can be \emph{thinned} down to a
Poisson process ${\Gamma_\omega \subset \Upsilon_\omega}$ with density $\varrho$  by 
deleting points $u\in {\Gamma'_\omega \subset \Upsilon_\omega}$ 
with probability $1/2$ and furthermore, we have that ${ \Gamma'_\omega = \Upsilon_\omega \setminus \Gamma_\omega}$ 
is also a Poisson process with density $\rho$ and ${\Gamma_\omega}$, ${ \Gamma'_\omega}$ 
are independent. Following \cite{germinetetal2005}, we  use
 this representation of ${\Gamma_\omega}$ to take care of the free sites.  For $B \sqcup S \in \mathcal {J}_\Lambda$, we 
define the \emph{$\Lambda$-bconfsets}, (definition 3.9 in \cite{germinetetal2005})
$$ C_{\Lambda,B,S} := \bigsqcup_{S' \subset S} [B \cup S ]_\Lambda,$$
and we define the \emph{$\Lambda$-bevents} (definition 3.10) as those $\omega$ such that,
for $B \sqcup B' \sqcup S \in \mathcal J_\Lambda$,  we have that $ \Gamma_\omega$ puts exactly one point in each $\Lambda(j)$ with $j \in B$, $ \Gamma'_\omega$ puts exactly one point in each $\Lambda(j)$ with $j \in B'$, and ${ \Upsilon}_\omega$ puts exactly one point in each $\Lambda(j)$ with $j \in S$ (so either $\Gamma_\omega$ or $\Gamma'_\omega$); and no points elsewhere, i.e.   
$$ C_{\Lambda, B,B',S} := \{  \Upsilon_\omega \in  [B \sqcup B' \sqcup S ]_\Lambda \} \cap \{\Gamma_\omega \in C_{\Lambda,B,S} \} \cap \{ \Gamma'_\omega \in C_{\Lambda,B',S}\}.$$

Now we proceed to the proof of the a priori estimate. We need to show that there exists a union of basic events inside which the resolvent decays exponentially, and that this union have good probability. As usual, once we know we are at a certain distance from the spectrum, the exponential decay is a consequence of the Combes--Thomas estimate.
Define $\mathcal {\hat J}_\Lambda$
$$\mathcal {\hat J}_\Lambda := \{ S \in \mathcal J_\Lambda: \, N_S(\Lambda_{\delta_L})(j) \leq 1 \textrm{ for all } j \in J \textrm{ and }  H_{B,\Lambda} \geq 2\varrho^{\alpha}\} . $$
As for any $\mathbf t_S \in [0,1]^S$ we have that
$$H_{B,(S,t_S),\Lambda} \geq H_{B,\Lambda}$$
we conclude that the set
$$\Omega_\Lambda := \bigsqcup_{(B,B',S) \in \hat {\mathcal J}_\Lambda} C_{\Lambda,B,B',S}$$
is $E$-localizing for $E \in [0,\varrho^{\alpha'}]$. Now, to prove that this happens with good probability, 
we see that if $B \in \mathcal P_0(\R^d)$ is such that
$$\inf \sigma(H_{B,\Lambda}) < 2 \varrho^{\alpha}$$
then for all $X \in [B]_\Lambda$, (see Lemma 3.8 in \cite{germinetetal2005})
$$H_{X,\Lambda} < C' \varrho^\alpha$$
and thus
$$\bigsqcup_{ {\mathcal J}_\Lambda \diagdown \hat {\mathcal J}_\Lambda} C_{\Lambda,B,B',S} \subset \{\omega: \inf \sigma(H_{\Gamma_\omega,(\phi,\phi),\Lambda}) < C \varrho^\alpha\}.$$
To estimate the probability of this set proceed as in the normalized Anderson case.

\subsection{Klopp's Periodic Approximations}

From now on we will take $N \in \N^*$ such that $(2N+1)$ is a multiple of $q$ (we will take  $q$ large but fixed for the Poisson potential).
Define the periodic approximation, for $\omega \in \Omega$ and 
\begin{align}
  H_{\omega}^N &= H_0 + \sum_{j \in \Z^d_{2N+1}} \omega_j \sum_{\zeta \in (2N+1)\Z^d} u(x-\zeta-j) \nonumber \\
  &= H_0 + V^N_\omega \nonumber 
\end{align}
for the normalized Anderson model and 
\begin{align}
  H_{\omega}^N &= H_0 + \sum_{\zeta \in (2N+1)\Z^d} \sum_{j \in {\Gamma_{\omega}^N}} u(x-\zeta-j) \nonumber \\
  &= H_0 + V_\omega^N \nonumber
\end{align}
for the Poisson--Anderson model, with $\Gamma_{\omega}^N=\Gamma_{\omega} \cap \Lambda_{N}(0)$. We write $\IDS_{\omega}^N$ the integrated density of states of this periodic operator.

From \cite{klopp2002b}, \cite{klopppastur1999}, we have the following:
\begin{lem}
 Let $\alpha>0$. There exists $\nu_0 \in (0,1)$ and $\gamma>0$ such that, for $\varrho \in [0,1]$, $E \in \R$, $\nu \in (0,\nu_0)$ and $N \geq \nu^{-\gamma}$ we have
\begin{displaymath}
 	\Esp(\IDS_{\omega}^N(E-\nu)) - e^{-\nu^{-\alpha}} \leq
 	\IDS(E) \leq
 	\Esp(\IDS_{\omega}^N(E+\nu)) + e^{-\nu^{-\alpha}}
\end{displaymath}
\end{lem}

As shown in section 2.3 in \cite{klopp2002b}, we estimate
\begin{align}
  \Esp \left( \IDS_\omega^N(E) \right) \leq C \Pro\left( \Omega(\varrho^\alpha, \varrho, N) \right) \nonumber 
\end{align}
where 
\begin{displaymath}
  \Omega(E,\varrho,N):=\{\omega: \sigma(H^N_{\omega}) \cap [0,E] \neq 0\}.
\end{displaymath}
or, by Floquet (see next section), we know that
\begin{equation}
\label{def:omega}
    \Omega(E,\varrho,N)=\{\omega:\exists ~ \theta \in \R^d \textrm{ s.t. }
	H_{\omega}^N (\theta) \textrm{ has an e.v. in } [0,E]\}.
\end{equation}

Theorem \ref{thm:lifschitzcontinuous} is thus a consequence of the following result.
\begin{prop}
\label{prop:contmain}
 Pick $\alpha> 2 \frac{d+1}{d}$ and $\gamma$ given by the last lemma. There exists $\varrho^*=\varrho^*(\alpha,\gamma) > 0$ and $\epsilon>0$ such that for $\varrho \in (0,\varrho^*)$ we have
\begin{displaymath}
 	\Pro[\Omega(\varrho^\alpha,\varrho,N)] \leq e^{-\varrho^{-\epsilon}}
\end{displaymath}
where
\begin{displaymath}
 	2N+1=q [\varrho^{(\alpha'-\alpha)/2}]_o [\varrho^{-\alpha'/4}]_o [\varrho^{-\gamma}]_o
\end{displaymath}
\end{prop}

\subsection{Floquet theory.}
We recall the corresponding Floquet theory for periodic operators
on the continuous.
For $\theta \in \T^d=\R^d /q \Z^d$, solving the problem:
\begin{displaymath}
\left\{ 
	\begin{array}{rclc}
		H_0\,\phi & = & \phi &\\
		\phi(x+j) & = & e^{i2\pi j \theta} \phi(x) &\textrm{; } (\forall x \in \R^d)(\forall j \in q\Z^d)
	\end{array} 
\right.
\end{displaymath}
yields Floquet eigenvalues $E_0(\theta) \leq \ldots E_n(\theta) \leq \ldots $ together with Floquet eigenvectors $\left( \phi_k(\theta) \right)_{k\geq 0}$. We recall also the following facts (\cite{klopp2002b}):

\begin{itemize}
  \item We write $\Sigma_0= \bigcup_{n\geq 0} E_n (\T^d)$, the spectrum of $H_0$.
  \item We have that the bottom of the spectrum is a simple non degenerate edge. This means that there exists $C>0$ such that:
  \begin{enumerate}
    \item[(P1)] For any $p>0$ and $\theta \in \T^d$, $\left| E_p(\theta) \right| \geq 1/C$ .
    \item[(P2)] There exists a set  $Z=\{ \theta_j;\, 1 \leq j \leq n_z \}$ such that
      $E_0(\theta_j)= \inf \Sigma_0=0$ and for $\theta  \in \T^d$,
      $$\left| E_0(\theta) \right| \geq C \min_{ 1 \leq j \leq n_z } |\theta-\theta_j|^2$$
  \end{enumerate}
  \item The density of states of $H_0$ satisfies (\cite{shubin1979spectral}):
\begin{displaymath}
  \IDS_0(E) = C_q \sum_{k\geq 1} \int_{\T^d} \Id_{E_k(\theta) \leq E}~d\theta.
\end{displaymath}
\end{itemize}

For $\theta \in \R^d$, let,
\begin{align}
  &\phantom{MMm} C^\infty_{N,\theta}(\R^d)=\left\{ \phi \in C^\infty(\R^d) \, \big| \, \phi(x+j)=e^{ij \theta} \phi(x) \textrm{; } j \in (2N+1)\Z^d \right\} \nonumber
\end{align}
and denote by $L^2_{N,\theta}(\R^d)$ (resp. $H^2_{N,\theta}(\R^d)$) the closure of this space in the $L^2_{\textrm{loc}}(\R^d)$ (resp. $H^2_{\textrm{loc}}(\R^d)$ Sobolev norm) norm, so
\begin{eqnarray}
    H^N(\theta):& \left\{ 
    \begin{array}{cccc}
  	&H^2_{N,\theta}(\R^d) &\to& L^2_{N,\theta}(\R^d) \\
  	&\phi &\mapsto& H\phi
    \end{array}
    \right.& \nonumber 
\end{eqnarray}

Now consider $H_0$ as a $(2N+1)\Z^d$-periodic operator, which we write $H_0^N$, and we write
\begin{align}
\left\{ 
	H_0^N(\theta)~:~
	\begin{array}{ccc}	
		H^2_{N,\theta}(\R^d) & \to & L^2_{N,\theta}(\R^d) \\
		\phi & \mapsto & H\phi
	\end{array} 
\right.
\end{align}
for its restriction to these spaces. We can verify that for $j \in \Z^d_{2N+1} = \Z^d/(2N+1)\Z^d$ and $\theta \in \T^d_{2N+1} = \R^d/\frac{1}{(2N+1)} \Z^d$, the Floquet eigenvalues and eigenvectors of $H_0^N(\theta)$:
\begin{displaymath}
  H_0 \phi_{k,j} (\cdot,\theta) = E_{k,j}(\theta) \phi_{k,j} (\cdot,\theta)
\end{displaymath}
where

\begin{displaymath}
  \left\{
	\begin{array}{ccc}
		E_{k,j}(\theta) &=& E_k (\theta + qj/(2N+1)) \\
		\phi_{k,j}(\cdot,\theta) &=& \frac{1}{(2 \pi + 1)^d} \phi_k (\cdot, \theta + \frac{qj}{(2N+1)}).
	\end{array}
  \right.
\end{displaymath}

Finally, for $\psi \in L^2(\R^d)$, we will use the decomposition:
\begin{eqnarray}
\label{eq:completefamily}
  \psi &=& \sum_{k \geq 0} \int_{\T^d} \widehat{\psi_k}(\theta)\phi_n(\cdot,\theta)~d\theta \\
  &=& \sum_{j \in \Z^d_{2N+1}} \sum_{k \geq 0} \int_{\T^d_{2N+1}} \widehat{\psi}_{j,k}(\theta)\phi_{j,k}(\cdot,\theta)~d\theta \nonumber.
\end{eqnarray}

\subsection{Proof of Proposition \ref{prop:contmain}}
The strategy of the proof of the Proposition follows the line of the proof of Proposition \ref{prop:main} and we will therefore omit some details (see also section 2.4 in \cite{klopp2002b}). 
Pick $\alpha>2 \frac{d+1}{d}$ , $\alpha'$ satisfing $\frac{d}{d+1} \alpha>\alpha'>2$ and large $\gamma$.
We define, as for the discrete case,
\begin{displaymath}
 2L+1=[\varrho^{(\alpha'-\alpha)/2}]_o[\varrho^{-\alpha'/4}]_o~~\textrm{et}~~2K+1=q [\varrho^{-\gamma}]_o. \nonumber 
\end{displaymath}

Let $\omega \in \Omega(\varrho^\alpha,\varrho,N)$. We have thus that there exists a  normalized $\psi \in H^2(\R^d)$ such that
\begin{equation}
  \left \langle H^N_{\omega} \psi,  \psi \right \rangle \leq \varrho^\alpha; \nonumber 
\end{equation}
by positivity, we also have
\begin{equation}
\label{eq:kinEcont}
  \left \langle  H_0^N \psi, \psi \right \rangle \leq \varrho^\alpha, 
\end{equation}
as well as
\begin{equation}
  \left \langle  V^N_{\omega} \psi, \psi \right \rangle \leq \varrho^\alpha.\nonumber
\end{equation}
Using \eqref{eq:kinEcont}, decomposition (\ref{eq:completefamily}) and (P1), (P2), we see that for $\psi \in H^2$ and $\varrho$ small enough, we know that,
\begin{equation}
  \sum_{k > 0} \int_{\T^d} |\widehat{\psi}_k(\theta)|^2 ~d\theta + \int_{\min |\theta-\theta_j|>1/L} |\widehat{\psi}_0(\theta)|^2 ~d\theta \lesssim \varrho^\alpha L^2, \nonumber 
\end{equation}
and decomposing
\begin{align}
  \psi=\sum_{1\leq j\leq n_z} \psi_j + \psi_e \nonumber 
\end{align}
with
\begin{align}
  \psi_j=\int_{|\theta-\theta_j| \leq 1/L} \widehat{\psi}_j(\theta) \phi_0(\theta) \, d\theta, \nonumber 
\end{align}
we have that, by the definition of $L$, $\left\| \psi_e \right\|_2 \lesssim \varrho^{\alpha'/4}$. 

As we did in the discrete setting, we expand
\begin{align}
  \left\langle V^N_\omega  \psi, \psi \right\rangle
    = \underbrace{\left\langle V^N_\omega    \sum_{1\leq j\leq n_z} \psi_j , \sum_{1\leq j\leq n_z} \psi_j  \right\rangle }_{(I)}
  + \underbrace{ 2 \, \textrm{Re} \Biggl\langle V^N_\omega  \psi_e,  \sum_{1\leq j\leq n_z} \psi_j \Biggl\rangle }_{(II)}
  + \underbrace{\Biggl\langle V^N_\omega  \psi_e, \psi_e \Biggl\rangle}_{(III)}  \nonumber
\end{align}
and similarly --- as we did after \eqref{eq:sum} --- we conclude, on the one hand, that  
$|(III)| \lesssim \varrho^{\alpha'/2}$, and, on the other, that if 
$\left| (II) \right| \gtrsim \varrho^{(6+\alpha')/8}$ 
we would have
\begin{align}
  \left\langle V^N_\omega   \sum_{1\leq j\leq n_z} \psi_j ,  \sum_{1\leq j\leq n_z} \psi_j  \right\rangle \gtrsim \varrho^{(6-\alpha')/4}  \nonumber 
\end{align}
or else
\begin{align}
   \left\langle V^N_\omega   \sum_{1\leq j\leq n_z} \psi_j ,  \sum_{1\leq j\leq n_z} \psi_j  \right\rangle \lesssim \varrho^{(6+\alpha')/8}. \nonumber
\end{align}
We now quote Lemma 2.1 in \cite{klopp2002b}, which says:
\begin{lem}
\label{lem:contapprox}
  Fix $1 \leq j \leq n_z$. For $1 \leq L' \leq L$, there exists $\tilde{\psi}_j \in L^2(\R^d)$ such that,
\begin{enumerate}
  \item The function $\tilde{\psi}_j$ is constant on each cube $\Lambda_{L'}(\gamma)$; $\gamma \in (2L'+1) \Z^d$. \label{enum:incert1}
  \item There exists $C>0$ such that
\begin{equation*}
  \| \psi_j(\cdot) - \tilde{\psi}_j(\cdot)\varphi_0(\cdot,\theta_j) \|_{2} \leq CL'/L
\end{equation*}
\label{enum:incert2}
where $\varphi_0(\cdot,\theta_j)$ is the periodic component $\phi_0(\cdot,\theta)$, i.e. 
\begin{align}
  \phi_0(\cdot,\theta)=e^{ix\theta} \varphi_0(\cdot,\theta).
\end{align} 
\end{enumerate}
\end{lem}
For the proof of this lemma we refer the reader to the end of the proof of Proposition 2.1 in 
\cite{klopp2002b}.

Let $2L'+1=[\varrho^{(\alpha'-\alpha)/2}]_o$ and $2K'+1=q[\varrho^{-\alpha'/4}]_o [\varrho^{-\gamma}]_o$.
By using the first point of Lemma \ref{lem:contapprox}, we write:
\begin{equation}
  \Psi_j(x)=\tilde{\psi}_j(x) \phi_0(x,\theta_j)=\phi_0(x,\theta_j) \sum_{\beta \in \Z^d} (2L'+1)^{-d/2} \alpha_j(\beta) \Id_{\Lambda_{L'}((2L'+1) \beta)'}(x), \nonumber 
\end{equation} 
and by posing $\Psi=\sum \Psi_j$ and $\alpha(\beta)=\sum \alpha_j(\beta)$ we have,
\begin{equation}
  \Psi(x)=\sum_j \tilde{\psi}_j(x) \phi_0(x,\theta_j)= \phi_0(x,\theta_j) \sum_{\beta \in \Z^d} (2L'+1)^{-d/2} \alpha(\beta) \Id_{\Lambda_{L'}({(2L'+1) \beta})}(x). \nonumber 
\end{equation} 
 Again, writing
\begin{align}
  \left \langle V^N_\omega  \sum_{1 \leq j\leq n_z} \psi_j, \sum_{1 \leq j\leq n_z}\psi_j \right \rangle  =  \left \langle V^N_\omega \sum_{1 \leq j\leq n_z}{\Psi}_j, \sum_{1 \leq j\leq n_z}{\Psi}_j \right \rangle  
  + & 2 \, \textrm{Re} \, \left \langle V^N_\omega \sum_{1 \leq j\leq n_z}{\Psi}_j, \sum_{1 \leq j\leq n_z}\left( \psi_j - {\Psi}_j \right) \right \rangle  \nonumber \\
  + &\left \langle V^N_\omega \sum_{1 \leq j\leq n_z}\left( \psi_j - {\Psi}_j \right), \sum_{1 \leq j\leq n_z}\left( \psi_j - {\Psi}_j \right) \right \rangle , \nonumber 
\end{align}
we see that, by the second point of Lemma \ref{lem:contapprox},
\begin{align}
  \left \langle V^N_\omega \sum_{1 \leq j\leq n_z}\left( \psi_j - {\Psi}_j \right), \sum_{1 \leq j\leq n_z}\left( \psi_j - {\Psi}_j \right) \right \rangle \lesssim \varrho^{\alpha'/2}, \nonumber 
\end{align}
and doing as in \eqref{eq:sum2} and thereafter, we conclude that 
\begin{align}
\label{ineq:contfinal}
    \pm \left \langle V^N_\omega {\Psi}, {\Psi} \right \rangle \lesssim \pm \varrho^{1 \pm \epsilon}.
\end{align}

We will now separate both cases. Consider first the Generalized Anderson model. 
For $k \in \Z^d_{\hat q}$, define:
$$V_{\omega,k}^N = \sum_{j \in \Z_{\frac{2N+1}{q}}^d} \omega_{qj+k} \sum_{\zeta \in (2N+1)\Z^d} u(x-\eta -qj -k)$$
so that
$$V_\omega^N = \sum_{k \in \Z^d_{\hat q}}V^N_{\omega,k}$$
and so the inequalities in $\eqref{ineq:contfinal}$ imply the same with $V^N_{\omega,k}$ instead of $V^N_{\omega}$, at least for one $k$ (and different constants). Note that $|\Z^d_{\hat q}|=q^d$ is finite and
independent of $\varrho$ so the probabilities, after the union bound,  will just change by a constant. 
As the calculation
is very similar for every $k$ we will assume that $k=0$ and we will drop it from the notation. Furthermore, we will assume that the support
of the simple site potential $u$ is entirely contained in the cell $\Lambda_{\hat q}(0)$, and we remember that $2 \hat q+1 = q$. If this is not the case we can change
$q$ by a multiple of $q$ large enough at the beginning of the analysis. 

We will denote from now on $\tilde V^N_\omega := V^N_\omega - \Esp[V_{per}]=V^N_\omega - \varrho V_{per}$ where $V_{per}$ is
the periodic operator which results if we take all random variables equal to $1$. A consequence of the
unique continuation principle, is that
\begin{align*}
\varrho \left\langle V_{per} \Psi,\Psi \right\rangle 
& = \varrho \left\langle V_{per} \psi,\psi \right\rangle + o(\varrho^{3/2})
\\ & \geq \varrho \left\langle V_{per} \psi,\psi \right\rangle +  \left\langle H_0 \psi,\psi \right\rangle -\varrho^\alpha + o(\varrho^{3/2})
\\ &=  \left\langle \left(H_0 + \varrho V_{per} \right) \psi,\psi \right\rangle + o(\varrho^{3/2}) \geq C \varrho
\end{align*}
and we obviously have $\varrho \left\langle V_{per} \Psi,\Psi \right\rangle \lesssim \varrho$.
\begin{rem}
Even without the unique continuation principle, the behavior of the bottom of the spectrum of the 
perturbed operator is of the order of the perturbation for a generic  simple site potential
$u$, as proven in \cite{klopp2002b}, section 5.
\end{rem} 
Using this, we conclude from \eqref{ineq:contfinal} that, there exists a $c$ such that for small $\varrho$
\begin{align}
\label{ineq:contfinal2}
     \left| \left \langle \tilde V^N_\omega {\Psi}, {\Psi} \right \rangle \right| \geq c\varrho.
\end{align}
We will show this happens with very low probability. As for every $j$ we have $\tilde{\psi}_j\in L^2(\R^d)$, let us calculate
\begin{align}
    \langle  \tilde V^N_\omega {\Psi}, {\Psi} \rangle &=  \sum_{\beta \in \Z^d} (2L'+1)^{-d}  \int_{\Lambda_{L'}({(2L'+1) \beta})}  \tilde V^N_\omega(x-j) \left| \sum_j \alpha_j(\beta)  \varphi_0(x,\theta_j) \right|^2 ~dx  \nonumber \\
    &= \sum_{\beta' \in \Z^d_{2K'+1}} \sum_{\beta'' \in (2K'+1)\Z^d} (2L'+1)^{-d}   \nonumber \\
     & \quad \int_{\Lambda_{L'}({(2L'+1) (\beta'+ \beta'')})}  \tilde V^N_\omega(x-j) \left| \sum_j \alpha_j(\beta'+\beta'')  \varphi_0(x,\theta_j)\right|^2  ~dx \nonumber
\end{align}
where, using the $(2N+1)$-periodicity of $ \tilde V^N_\omega$ and the fact that $(2L'+1)(2K'+1)=2N+1$, the last line is equal to
\begin{align}
\nonumber &   \sum_{\beta' \in \Z^d_{2K'+1}} (2L'+1)^{-d} \sum_{\beta'' \in (2K'+1)\Z^d}  \int_{\Lambda_{L'}((2L'+1) \beta')} \tilde V^N_\omega(x-j) \left|  \sum_j \alpha_j(\beta'+\beta'')\varphi_0(x,\theta_j) \right|^2 ~dx \\ 
\nonumber &=  \sum_{\beta''' \in \Z^d_{\frac{2L'+1}{q}}}  \sum_{\beta' \in \Z^d_{2K'+1}} (2L'+1)^{-d}  \int_{\Lambda_{\hat q}({(2L'+1) \beta'+q\beta'''})}  \tilde V^N_\omega(x-j) \sum_{\beta'' \in (2K'+1)\Z^d} \left|  \sum_j \alpha_j(\beta'+\beta'')\varphi_0(x,\theta_j) \right|^2 ~dx \\
\nonumber &= \frac{\sum_{\beta''' \in \Z^d_{2L'+1}} X(\beta''')}{(2L'+1)^d}.
\end{align}
The random variables $X(\beta''')$ are independent, bounded, non trivial and their expectation $\mathbb E\left[X(\beta''')\right] =0$. As usual we will prove only one side of the large deviation inequality.
Reindex the random variables as $X_U$, $U=1,\ldots,R=(2L'+1)^d$; then use Markov's inequality to obtain:
\begin{align}
  \Pro\left( \frac{1}{R} \sum_{U=1}^R X_U \geq c \varrho \right) \leq \Esp\left( e^{ t \sum X_U} \right) e^{-cRt\varrho} \\
  \leq \prod_{U=1}^R \Esp\left( e^{ t X_U} \right) e^{ -CRt\varrho}.
  \nonumber
\end{align}
Now if we take $t$ small enough,
thus
\begin{align}
  \nonumber \Esp\left( e^{t \omega_U} \right) &\leq e^{ c t^2 \Esp(X_U^2)},
\end{align}
and thus, noting that $\Esp(X_U ^2) \lesssim \varrho$,
\begin{align}
\nonumber
  \Pro\left( \frac{1}{R} \sum_{U=1}^R X_U \geq c \varrho \right)    \leq  e^{ c R t^2 \varrho -CRt\varrho} \leq e^{-c'R\varrho}.
\end{align}
Here $R \sim \rho^{-1-\epsilon'}$;  this probability is exponentially decaying. This --and summing up all the probabilities--  proves what we wanted for the normalized Anderson model..

We turn our attention now to  the Poisson--Anderson model. We define $V^N_{\omega,k}$ in a similar way:
$$V^N_{\omega,k}:= \sum_{\zeta \in (2N+1)\Z^d} \sum_{j \in \Gamma_{\omega,k}} u(x-\xi-j)$$
where we have defined 
$$\Gamma_{\omega,k}:=\Gamma_{\omega} \cap \left( \cup_{n \in Z^d_{\frac{2N+1}{q}}} \Lambda_0({nq+k}) \right) $$ 
for $k \in \Z^d_{\hat q}$. (Note that $\Lambda_0(\cdot)$ is a unit cube.) We have thus the equality $V^N_\omega = \sum V^N_{\omega,k}$ with 
each $V^N_{\omega,k}$ positive. Inequalities \eqref{ineq:contfinal} lead to the same inequalities
with $V^N_\omega$ replaced by $V^N_{\omega,k}$, for at least one $k \in \Z^d_{\hat q}$. We suppose
as before that $k=0$ and we drop it from the notation, the others being similar. Again, the probability
will be bounded by the union bound on a finite number of events. 

As here $H_0=\triangle$, there is only one minimum
of the Floquet eigenvalue at $0$, and $\phi_0(x,0)$ is a constant function. Define the random variable
\begin{align}
    \chi(\beta,R)=\# \left[ \left( \bigcup_{n \in (2N+1) \Z^d}  \Gamma_\omega \cap \Lambda_R(n+\beta) \right)  \cap \Lambda_N(0)\right] \nonumber 
\end{align}
so
\begin{align}
    \langle  V^N_\omega {\Psi}, {\Psi} \rangle &=  \sum_{\beta \in  \Z^d} (2L'+1)^{-d}   \left| \alpha_0(\beta)  \right|^2  \int_{\Lambda_{L'}({(2L'+1) \beta})} V^N_\omega(x-j)~dx  \nonumber \\
    &=  \sum_{\beta''' \in Z^d_{\frac{2L'+1}{q}}} \sum_{\beta \in \Z^d} (2L'+1)^{-d} \left| \alpha_0(\beta)  \right|^2 \int_{\Lambda_{\hat q}((2L'+1) \beta+q\beta''')} V^N_\omega(x-j)  ~dx  \nonumber \\
    &  = c \sum_{\beta''' \in Z^d_{\frac{2L'+1}{q}}} \sum_{\beta \in \Z^d} (2L'+1)^{-d} \left|  \alpha_0(\beta)  \right|^2 \chi\left( (2L'+1)\beta + q\beta''',0 \right). \nonumber 
\end{align}
So  \eqref{ineq:contfinal}, for $\varrho$ small enough, becomes
\begin{align}
\pm \sum_{\beta''' \in Z^d_{2L'+1}} \frac{X(\beta''')}{(2L'+1)} \lesssim \pm \varrho^{1\pm\epsilon}, \nonumber
\end{align}
with $X(\beta''')=  c \sum_{\beta \in \Z^d}  \left|  \alpha_0(\beta)  \right|^2 \chi\left( (2L'+1)\beta + \beta''',0 \right)$. Note that we have chosen $q$ large enough --but independent of $\varrho$-- so these random variables are independent. This probability can be  again estimated by a large deviation type estimate to get the desired result.

\addcontentsline{toc}{section}{Bibliography}
\bibliographystyle{plain}
\bibliography{biblio-lowdens}

\end{document}

%% file: preambule-english.tex
\usepackage{amssymb,bbm}
\usepackage{amsmath}
\usepackage{amsmath}
\usepackage{amsfonts}
\usepackage{amssymb}
\usepackage[latin1]{inputenc}
\usepackage[english]{babel}
\usepackage{makeidx}
\usepackage{index}
\usepackage{empheq}
\usepackage{txfonts}
\usepackage{hyperref}
\usepackage{url}
\usepackage{index}

\DeclareMathOperator*{\supp}{supp}

\newcommand{\Id}{\mathbf{1}}

\newcommand{\Hi}{\mathcal{H}}

\newcommand{\IDS}{\mathcal{N}}
\newcommand{\Fourier}{\mathcal{F}}
\newcommand{\Pro}{\mathbb{P}}
\newcommand{\Esp}{\mathbb{E}}
\newcommand{\R}{\mathbb{R}}

\newcommand{\N}{\mathbb{N}}
\newcommand{\T}{\mathbb{T}}
\newcommand{\Z}{\mathbb{Z}}

\newcommand{\findem}{\begin{flushright} $\blacksquare$ \end{flushright}}

\newtheorem{thm}{Theorem}[section]
\newtheorem{prop}[thm]{Proposition}

\newtheorem{lem}[thm]{Lemma}
\newtheorem{rem}[thm]{Remark}